\documentclass[showpacs]{revtex4} 
\usepackage{epsfig,amsmath,amssymb,graphicx,upgreek,textcomp}
\usepackage{natbib}
\usepackage[english]{babel} %

\begin{document}

\vspace{0mm}
\title{On the thermodynamics of two-level Fermi and Bose nanosystems } %
\author{Yu.M. Poluektov}
\email{yuripoluektov@kipt.kharkov.ua (y.poluekt52@gmail.com)} %
\affiliation{National Science Center ``Kharkov Institute of Physics and Technology'', 61108 Kharkov, Ukraine} %
\author{A.A. Soroka} %
\affiliation{National Science Center ``Kharkov Institute of Physics and Technology'', 61108 Kharkov, Ukraine} %

\begin{abstract}
Equations are obtained for the quantum distribution functions over
discrete states in systems of non-interacting fermions and bosons
with an arbitrary, including small, number of particles. The case of
systems with two levels is considered in detail. The temperature
dependences of entropy, heat capacities and pressure in two-level
Fermi and Bose systems are calculated for various multiplicities of
degeneracy of levels.
\newline%
\end{abstract}
\maketitle

\section{Introduction}\vspace{-0mm} 
Currently, much attention is paid to the study of quantum properties
of systems with a small number of particles, such as quantum dots
and other mesoscopic objects and nanostructures. In this regard, the
problem of describing the properties of such objects with taking
into account their interaction with the external environment is actual. %

Statistical description is usually used to study systems with a very
large number of particles $N$ in a large volume $V$ with a
subsequent transition to the thermodynamic limit %
$N\rightarrow\infty, V\rightarrow\infty$ at $n=N/V=const$. %
However, statistical methods can also be applied to the study of
equilibrium states of systems with a small number of particles and
even one particle in a finite volume. When considering a
many-particle system within the framework of a grand canonical
ensemble, it is assumed that it is a part of a very large system, a
thermostat, with which it can exchange energy and particles. The
thermostat itself is characterized by such statistical quantities as
temperature $T$ and chemical potential $\mu$. Assuming that the
subsystem under consideration is in thermodynamic equilibrium with
the thermostat, the subsystem itself is characterized by the same
quantities, even one consisting of a small number of particles. For
example, we can consider the thermodynamics of an individual quantum
oscillator [1]. In the case when an exchange of particles with a
thermostat is possible, the time-averaged number of particles of a
small subsystem may be not an integer and may even be less than
unity. For this case, the equations that determine the average
number of particles in each state for the Fermi-Dirac and
Bose-Einstein statistics were obtained by the authors in [2]. In
this work, based on the theory proposed in [2], the temperature
dependences of entropy, heat capacities and pressure in systems with
two levels are calculated.

The model of quantum objects with two states is used to describe a
wide range of phenomena [3,4]. The concept of two-level systems was
initiated by the phenomena of magnetic resonance and was further
developed in connection with the advent of lasers. Issues related to
the equilibrium thermodynamics of two-level systems turned out to be
less studied [4]. The two-level model is also applicable for
describing multilevel systems at temperatures lower than the energy
difference between the second and the next after it energy level.

The second section of the article presents general relations for
entropy, distribution functions of particles and thermodynamic
quantities of an arbitrary number of fermions and bosons for a
system with an arbitrary number of levels. In the third and fourth
sections the two-level systems of bosons and fermions are studied in
detail, the temperature dependences of their entropy, heat
capacities, pressure and populations of levels are calculated. In
conclusion the obtained results are discussed.

\section{Entropy and distribution functions of fermions and bosons}\vspace{-0mm} %
In this section we present a brief derivation of equations for the
average numbers of particles in quantum states and formulas for
thermodynamic quantities for an arbitrary number of levels and
number of particles, which are valid for both Bose-Einstein and
Fermi-Dirac statistics [2]. Let us consider a quantum system of
non-interacting particles, the energy levels $\varepsilon_j$ of
which have the multiplicities of degeneracy $z_j$. If there are
$N_j$ particles at each level $j$, then $n_j=N_j\big/z_j$ is the
average number of particles at level $j$ or the population of the
level. In the case when for fermions $n_j\neq 0,1$ and for bosons
$n_j\neq 0$, the average number of particles in each state is found
from the condition of an extremum of the entropy $S=\sum_jS_j$
\begin{equation} \label{01}
\begin{array}{l}
\displaystyle{%
   \frac{\partial}{\partial n_j}\big(S-\alpha N-\beta E\big)=0,  %
}
\end{array}
\end{equation}
where the total number of particles $N$ and the total energy $E$ are
determined by the formulas
\begin{equation} \label{02}
\begin{array}{l}
\displaystyle{%
   N=\sum_jN_j=\sum_jn_jz_j,  %
}
\end{array}
\end{equation} \vspace{-5mm}
\begin{equation} \label{03}
\begin{array}{l}
\displaystyle{%
   E=\sum_j\varepsilon_jN_j=\sum_j\varepsilon_jn_jz_j.  %
}
\end{array}
\end{equation}
In (1) $\alpha,\beta$ are the Lagrange multipliers, which are found
from a comparison with thermodynamic relations, wherefrom it follows
$\alpha=-\mu/T,\, \beta=1/T$, $T$ -- temperature, $\mu$ -- chemical
potential [1]. Let us introduce the notation for the derivative of entropy: %
\begin{equation} \label{04}
\begin{array}{l}
\displaystyle{%
   \frac{\partial S_j}{\partial n_j}\equiv z_j\,\theta_j(n_j).  %
}
\end{array}
\end{equation}
The expression for entropy $S_j$ and the form of functions
$\theta_j(n_j)$ are different for fermions and bosons, and will be
obtained in the following sections. From (1)\,--\,(4) we find the
equations that determine the population of levels
\begin{equation} \label{05}
\begin{array}{l}
\displaystyle{%
   \theta_j(n_j)=\frac{(\varepsilon_j-\mu)}{T}.  %
}
\end{array}
\end{equation}
For a fixed total number of particles, due to condition (2), these
equations are not independent.

When constructing the thermodynamics of systems located in a limited
volume, the dependence of the level energy on volume should be taken
into account. This dependence arises as a consequence of the
boundary condition for the wave function at the boundary of the
volume $V$. For both the sphere and the cube $\varepsilon_j\sim V^{\,-2\!/3}$. %
Therefore, we will assume that $\varepsilon_j=\xi_jV^{\,-\alpha}$, %
where $\alpha>0$, so that $d\varepsilon_j\big/dV=-\alpha\big(\varepsilon_j/V\big)$, %
$d^{\,2}\varepsilon_j\big/dV^2=\alpha(\alpha+1)\big(\varepsilon_j/V^2\big)$. %
In numerical calculations we always assume $\alpha=2/3$. For the
thermodynamic potential $\Omega=E-TS-\mu N$ the known relation
$d\Omega=-SdT-Nd\mu-pdV$ is valid, therefore the pressure is
determined from the condition
\begin{equation} \label{06}
\begin{array}{l}
\displaystyle{%
   p=-\bigg(\frac{\partial\Omega}{\partial V}\bigg)_{\!T,\mu}=-\sum_jz_jn_j\frac{d\varepsilon_j}{dV}=\frac{\alpha}{V}\sum_jz_jn_j\varepsilon_j.  %
}%
\end{array}
\end{equation}

To calculate heat capacities and thermodynamic coefficients it is
necessary to use the expression for the differential of population,
which follows from (5):
\begin{equation} \label{07}
\begin{array}{l}
\displaystyle{%
  z_j\theta_j^{(1)}(n_j)\,dn_j =-\theta_j(n_j)\frac{dT}{T}-\frac{d\mu}{T}+\frac{1}{T}\frac{d\varepsilon_j}{dV}\,dV,  %
}%
\end{array}
\end{equation}
where
\begin{equation} \label{08}
\begin{array}{l}
\displaystyle{%
   \frac{\partial \theta_j(n_j)}{\partial n_j}\equiv z_j\,\theta_j^{(1)}(n_j).  %
}
\end{array}
\end{equation}
With taking into account (7),\,(8) we find the differentials of the
number of particles $N$, entropy $S$ and pressure $p$\,:
\begin{equation} \label{09}
\begin{array}{l}
\displaystyle{%
   dN=-A_1\frac{dT}{T}-A\frac{d\mu}{T}+B\frac{dV}{T}, %
}
\end{array}
\end{equation} \vspace{-4mm}
\begin{equation} \label{10}
\begin{array}{l}
\displaystyle{%
   dS=-A_2\frac{dT}{T}-A_1\frac{d\mu}{T}+B_1\frac{dV}{T},  %
}
\end{array}
\end{equation} \vspace{-4mm}
\begin{equation} \label{11}
\begin{array}{l}
\displaystyle{%
   dp=B_1\frac{dT}{T}+B\frac{d\mu}{T}-DdV.  %
}
\end{array}
\end{equation}
In (9)\,--\,(11) the following notations are used:
\begin{equation} \label{12}
\begin{array}{c}
\displaystyle{%
  A\equiv\sum_j\frac{1}{\theta_j^{(1)}}, \quad A_1\equiv\sum_j\frac{\theta_j}{\theta_j^{(1)}}, \quad A_2\equiv\sum_j\frac{\theta_j^2}{\theta_j^{(1)}},  %
}\vspace{3mm}\\ %
\displaystyle{\hspace{0mm}%
 B\equiv\sum_j\frac{1}{\theta_j^{(1)}}\frac{d\varepsilon_j}{dV}=-\frac{\alpha}{V}\sum_j\frac{\varepsilon_j}{\theta_j^{(1)}}, \quad %
 B_1\equiv\sum_j\frac{\theta_j}{\theta_j^{(1)}}\frac{d\varepsilon_j}{dV}=-\frac{\alpha}{V}\sum_j\frac{\theta_j\varepsilon_j}{\theta_j^{(1)}}, %
}\vspace{3mm}\\ %
\displaystyle{\hspace{0mm}%
 D\equiv\sum_j\!\Bigg[z_jn_j\frac{d^{\,2}\varepsilon_j}{dV^2}+\frac{1}{T\theta_j^{(1)}}\bigg(\frac{d\varepsilon_j}{dV}\bigg)^{\!2}\Bigg]= %
 \frac{\alpha}{V^2}\sum_j\!\Bigg[(1+\alpha)z_jn_j\varepsilon_j+\frac{\alpha\varepsilon_j^2}{T\theta_j^{(1)}}\Bigg]. %
}%
\end{array}
\end{equation}
Systems with a fixed number of particles are usually considered. In
this case $dN=0$, and the differential of the chemical potential can
be excluded from (9)\,--\,(11). As a result, we arrive at the
following formulas for the isochoric and isobaric heat capacities: %
\begin{equation} \label{13}
\begin{array}{l}
\displaystyle{%
   C_{V,N}=\frac{\big(A_1^2-AA_2\big)}{A},  %
}
\end{array}
\end{equation} \vspace{-5mm}
\begin{equation} \label{14}
\begin{array}{l}
\displaystyle{%
  C_{p,N}=\frac{\big(A_1^2-AA_2\big)}{A}-\frac{\big(AB_1-A_1\!B\big)^2}{A\big(B^2-ADT\big)}\,. %
}
\end{array}
\end{equation}
We also present formulas for the coefficient of volumetric expansion
\begin{equation} \label{15}
\begin{array}{l}
\displaystyle{%
  \alpha_{pN}=\frac{1}{V}\bigg(\frac{\partial V}{\partial T}\bigg)_{\!p,N}=-\frac{\big(AB_1-A_1\!B\big)}{V\big(B^2-ADT\big)},  %
}
\end{array}
\end{equation}
the isothermal compressibility
\begin{equation} \label{16}
\begin{array}{l}
\displaystyle{%
  \gamma_{T\!N}=-\frac{1}{V}\bigg(\frac{\partial V}{\partial p}\bigg)_{\!T,N}=-\frac{AT}{V\big(B^2-ADT\big)}  %
}
\end{array}
\end{equation}
and the isochoric thermal pressure coefficient
\begin{equation} \label{17}
\begin{array}{l}
\displaystyle{%
  \beta_{V\!N}=\frac{1}{p}\bigg(\frac{\partial p}{\partial T}\bigg)_{\!V,N}=\frac{\big(AB_1-A_1\!B\big)}{p\,TA}.  %
}
\end{array}
\end{equation}
The remaining thermodynamic coefficients can be found from the above
coefficients and the heat capacities [6]. The conditions for
thermodynamic stability of a system are the inequalities %
$\big(\partial p/\partial V\big)_T<0$ and $C_{V,N}>0$ [1]. %
We also note that the general thermodynamic relation %
$C_p-C_V=TV\big(\alpha_p^2\,/\gamma_T\big)$ turns out to be valid.

\section{Two-level boson system}\vspace{-0mm} %
Using the general formulas given in the previous section, we
consider a special case of a system of bosons which can exist only
in two states with energies $\varepsilon_2>\varepsilon_1$ and
multiplicities of degeneracy $z_1$ and $z_2$. If in a system of
bosons at each level with the multiplicity of degeneracy $z_j$ there
are $N_j$ particles, then the statistical weight of such a state in
the Bose-Einstein statistics [1]
\begin{equation} \label{18}
\begin{array}{l}
\displaystyle{%
   \Delta\Gamma_j = \frac{(z_j+N_j-1)!}{N_j!\,(z_j-1)!}. %
}
\end{array}
\end{equation}
In the general case, when the number of particles can be small and
take fractional values, the factorials in (18) should be determined
through the gamma function $N!=\Gamma\big(N+1\big)$ [2], so that the
statistical weight (18) will be written in the form
\begin{equation} \label{19}
\begin{array}{l}
\displaystyle{%
   \Delta\Gamma_j = \frac{\Gamma(z_j+N_j)}{\Gamma(N_j+1)\,\Gamma(z_j)}. %
}
\end{array}
\end{equation}
This implies the formula for the nonequilibrium entropy
$S=\sum_jS_j=\sum_j\ln\Delta\Gamma_j$, where
\begin{equation} \label{20}
\begin{array}{l}
\displaystyle{%
   S_j\equiv S_{B\!j}=\ln\Gamma\big(z_jn_j+z_j\big) - \ln\Gamma\big(z_jn_j+1\big) - \ln\Gamma\big(z_j\big). %
}
\end{array}
\end{equation}
In this case, in the equations of the previous section that
determine the population of levels and thermodynamic quantities, one
should use the formulas
\begin{equation} \label{21}
\begin{array}{l}
\displaystyle{%
   \theta_j(n_j)\equiv\theta_{B\!j}(n_j)\equiv\psi\big(z_jn_j+z_j\big) - \psi\big(z_jn_j+1\big), %
}
\end{array}
\end{equation}\vspace{-5mm}
\begin{equation} \label{22}
\begin{array}{l}
\displaystyle{%
   \theta_j^{(1)}(n_j)\equiv\theta_{B\!j}^{(1)}(n_j)\equiv\psi^{(1)}\big(z_jn_j+z_j\big) - \psi^{(1)}\big(z_jn_j+1\big), %
}
\end{array}
\end{equation}
where $\psi(x)\equiv d\ln\Gamma(x)\big/dx,\, \psi^{(1)}(x)\equiv
d^{\,2}\ln\Gamma(x)\big/dx^2$ are the logarithmic derivatives of the
gamma function [5]. Functions (21),\,(22) can be calculated using the formulas %
\begin{equation} \label{23}
\begin{array}{l}
\displaystyle{%
  \theta_{B\!j}(n_j)=\int_0^\infty \frac{e^{-z_jn_jt}\big(1-e^{-(z_j-1)t}\big)}{e^t-1}\,dt, \quad %
  \theta_{B\!j}^{(1)}(n_j)=-\int_0^\infty t\,\frac{e^{-z_jn_jt}\big(1-e^{-(z_j-1)t}\big)}{e^t-1}\,dt. %
}%
\end{array}
\end{equation}

At zero temperature all bosons are located at the ground level, so
that in this case $\mu=\varepsilon_1$ and the total number of
particles, energy and pressure are given by formulas $N=z_1n_1$,
$E=\varepsilon_1 N$ and $p=\alpha\big(N/V\big)\varepsilon_1$. The
entropy, when taking into account the discreteness of levels, at
$T=0$ turns out to be different from zero
\begin{equation} \label{24}
\begin{array}{l}
\displaystyle{%
   S_{B0}=\ln\Gamma\big(N+z_1\big) - \ln\Gamma\big(N+1\big) - \ln\Gamma\big(z_1\big). %
}
\end{array}
\end{equation}
Thus, the third law of thermodynamics is satisfied in the Nernst
formulation, according to which all processes at zero temperature
occur at a constant entropy, but not in the Planck formulation which
requires turning of the entropy to zero.

\vspace{0mm} %
\begin{figure}[b!]
\vspace{0mm}  \hspace{0mm}
\includegraphics[width = 7.4cm]{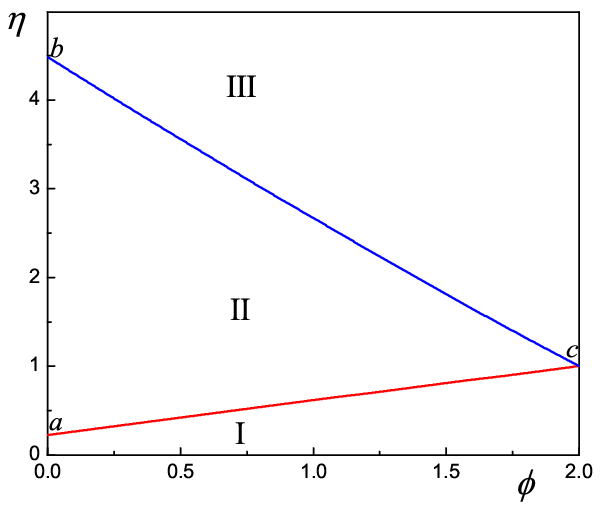} 
\vspace{-4mm} %
\caption{\label{fig01} 
Regions of existence of the characteristic temperatures (25),\,(26)
on the plane $\big(\phi,\eta\big)$, for $N=2$: I -- $T_{B1}, T_{B2}$
do not exist; II -- only $T_{B1}$ exists; III -- both temperatures $T_{B1}, T_{B2}$ exist. %
On the curve {\it ac}\, $\Phi_B(0,\phi)=0$, on the curve {\it bc}\, $\Phi_B(\phi/\eta,0)=0$. %
Coordinates of points: {\it a} -- (0,\,0.22); {\it b} -- (0,\,4.5); {\it c} -- (2,\,1). %
}%
\end{figure}

In the two-level Bose system, two characteristic temperatures can be
defined. At one of them the transition of particles from the lower
level to the upper level begins
\begin{equation} \label{25}
\begin{array}{l}
\displaystyle{%
   T_{B1}=\frac{\Delta\varepsilon}{\Phi_B\big(0,N/z_1\big)}, %
}
\end{array}
\end{equation}
and at the second characteristic temperature all particles transit
from the lower level to the upper level
\begin{equation} \label{26}
\begin{array}{l}
\displaystyle{%
   T_{B2}=\frac{\Delta\varepsilon}{\Phi_B\big(N/z_2,0\big)}, %
}
\end{array}
\end{equation}
where $\Delta\varepsilon\equiv\varepsilon_2-\varepsilon_1 > 0$. In
(25),\,(26) the notation is also used %
$\Phi_B(n_2,n_1)\equiv\theta_{B2}(n_2)-\theta_{B1}(n_1)$. Obviously,
the condition of the existence of these temperatures is the
positivity of denominators in (25),\,(26). These requirements are
satisfied not for all values of the quantities $z_1,z_2,N$. There
are three possibilities: I) the existence condition is not satisfied
in both cases, and the temperatures $T_{B1}, T_{B2}$ do not exist;
II) the existence condition is satisfied for (25), but it is not
satisfied for (26), so only the temperature $T_{B1}$ exists and the
temperature $T_{B2}$ is missing; III) both temperatures $T_{B1},
T_{B2}$ exist. It is convenient to introduce variables:
\begin{equation} \label{27}
\begin{array}{l}
\displaystyle{%
   \eta\equiv\frac{z_2}{z_1}, \qquad \phi\equiv\frac{N}{z_1}. %
}
\end{array}
\end{equation}
The regions I, II, III on the plane $\big(\phi,\eta\big)$ are shown
in Fig.\,1. In the figures, for convenience, we assume that $\eta$
and $\phi$ can take arbitrary positive values. Only those values of
$\eta$ and $\phi$ correspond to the physical parameters, for which
conditions (27) are satisfied. In the notation (27) %
$T_{B1}=\Delta\varepsilon\big/\Phi_B(0,\phi)$ and %
$T_{B2}=\Delta\varepsilon\big/\Phi_B(\phi\eta^{-1},0)$. %
Note that functions $\Phi_B(0,\phi)$ and $\Phi_B(\phi\eta^{-1},0)$
depend also on the number of particles $N$ as on a parameter.

The dependences of the dimensionless temperatures
$\tau_{B1}=T_{B1}\big/\Delta\varepsilon$ and $\tau_{B2}=T_{B2}\big/\Delta\varepsilon$ %
on $\eta$ for a fixed parameter $\phi$ are shown in Fig.\,2. In the
region I (Fig.\,1) the temperatures $T_{B1}, T_{B2}$ do not exist.
Physically, this means that in the case when the multiplicity of
degeneracy of the upper level is considerably less than the
multiplicity of degeneracy of the ground level ($\eta<\eta_1$ in
Fig.\,2), all particles remain locked at the ground level at
arbitrary permissible temperatures, and $n_2=0$, so that energy
cannot be transferred to the system and it remains adiabatically
isolated from the thermostat. In this case the pressure and entropy
remain constant, and the heat capacities are equal to zero.
\vspace{0mm} %
\begin{figure}[h!]
\vspace{-3mm}  \hspace{0mm}
\includegraphics[width = 7.4cm]{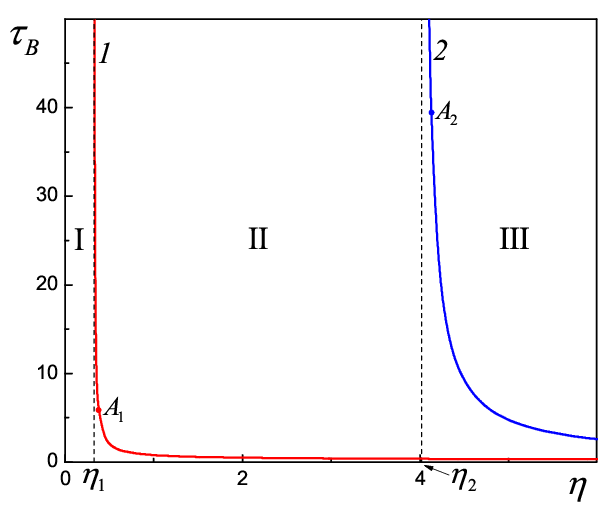} 
\vspace{-5mm} %
\caption{\label{fig02} 
Dependences of the temperatures  $\tau_{B1}$ \!({\it 1}) and $\tau_{B2}$ ({\it 2}) on $\eta$ %
for $z_1=8$, $N=2$, $\phi=0.25$; the point $A_1=(0.38,\,5.85)$ for $z_2=3$, %
the point $A_2=(4.13,\,39.44)$ for $z_2=33$; $\eta_1=0.33$, $\eta_2=4.02$. %
}%
\end{figure}

In the region II (Fig.\,1) there is the temperature $T_{B1}$
($\eta_1<\eta<\eta_2$ in Fig.\,2) at which the transition of
particles from the lower to the upper level begins, and the
dependence of populations on temperature at $T>T_{B1}$ is determined
by the system of equations
\begin{equation} \label{28}
\begin{array}{l}
\displaystyle{%
   \theta_{B1}(n_1)=\frac{(\varepsilon_1-\mu)}{T}, \qquad \theta_{B2}(n_2)=\frac{(\varepsilon_2-\mu)}{T}, \qquad %
   N=z_1n_1 + z_2n_2.
}
\end{array}
\end{equation}

In the limit of high temperatures $T\rightarrow\infty$ the
populations of levels tend to constant values $n_{1\infty}$ and
$n_{2\infty}$, which are determined by the conditions
$\theta_{B1}(n_{1\infty})=\theta_{B2}(n_{2\infty})$ and %
$N=z_1n_{1\infty} + z_2n_{2\infty}$. The temperature dependences of
the entropy and heat capacities for this case are shown in Fig.\,3{\it a}. %
At temperatures $T<T_{B1}$ the entropy $S_{B1}$ and pressure are
constant, and the heat capacities are zero. At temperature $T_{B1}$
the entropy and pressure begin to increase monotonically, and the
heat capacities take on by jumps finite values. Initially, the heat
capacities increase reaching their maxima, and with a further
increase in temperature they decrease monotonically. In the limit
$T\rightarrow\infty$ the entropy tends to a finite value
$S_{B\infty}$, and the heat capacities tend to zero.

\vspace{0mm} %
\begin{figure}[t!]
\vspace{-2mm}  \hspace{0mm}
\includegraphics[width = 15cm]{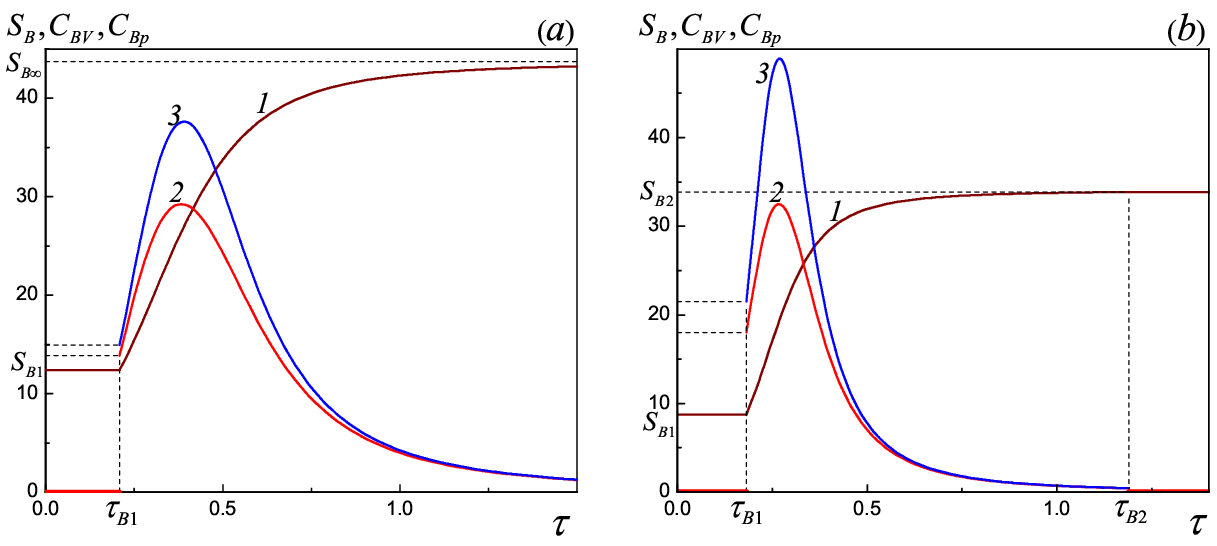} 
\vspace{-4mm} %
\caption{\label{fig03} 
Dependences of the entropy $S_B(\tau)$ ({\it 1}), the heat
capacities $C_{BV}(\tau)$ ({\it 2}) and $C_{Bp}(\tau)$ ({\it 3}) on
the dimensionless temperature $\tau=T/\Delta\varepsilon$ for the two-level boson system. %
(\!{\it a}) Region II. Jumps of heat capacities at $\tau_{B1}$: $\Delta C_{BV}=13.84$, $\Delta C_{Bp}=14.92$. %
Entropy values: $S_{B1}=12.41$, $S_{B\infty}=43.74$. %
Parameters: $z_1=8, z_2=96, N=16$; $\tau_{B1}=0.21$,
$n_{1\infty}=0.08$, $n_{2\infty}=0.16$. \newline %
(\!{\it b}) Region III. Jumps of heat capacities at $\tau_{B1}$: $\Delta C_{BV}=17.99$, $\Delta C_{Bp}=21.49$; %
jumps at $\tau_{B2}$: $\Delta C_{BV}=-0.43$, $\Delta C_{Bp}=-0.44$. %
Entropy values: $S_{B1}=8.77$, $S_{B2}=33.87$. %
Parameters: $z_1=8$, $z_2=256$, $N=8$; $\tau_{B1}=0.18$, $\tau_{B2}=1.2$. %
}%
\vspace{-3mm}
\end{figure}

In the region III (Fig.\,1) there exist two temperatures $T_{B1},
T_{B2}$ ($\eta>\eta_2$ in Fig.\,2). In this case at $T_{B1}$ the
transition of particles from the lower to the upper level begins,
and at $T=T_{B2}$ all particles transit to the upper level, so that
the lower level proves to be empty $n_1=0$. In the temperature range
$T_{B1}<T<T_{B2}$ the temperature dependence of populations is
determined by equations (28). The temperature dependences of the
entropy and heat capacities for this case are shown in Fig.\,3{\it b}. %
Here also at temperatures $T<T_{B1}$ the entropy $S_{B1}$ is
constant and the heat capacities are zero. At temperature $T_{B1}$
the entropy begins to increase monotonically up to a maximum value
$S_{B2}$ at $T_{B2}$, and the heat capacities take on by jumps
finite values. With an increase in temperature the heat capacities
first increase, and then they begin to decrease down to finite
values at temperature $T_{B2}$. At $T>T_{B2}$ the entropy $S_{B2}$
and pressure remain constant, and the heat capacities turn to zero
by jumps. In this region of high temperatures the energy of the
system reaches its maximally possible value and no longer increases
with increasing the thermostat temperature.

\section{Two-level fermion system}\vspace{-0mm} %
Let us consider a system of fermions which can exist in two states
with energies $\varepsilon_2>\varepsilon_1$ and multiplicities of
degeneracy $z_1$ and $z_2$. Such a system can contain no more than
$z_1+z_2$ particles. If there are $N_j\leq z_j$ particles at each
level $j$, then the statistical weight of such a state in the case
of Fermi-Dirac statistics is given by the well-known formula [1] %
\begin{equation} \label{29}
\begin{array}{l}
\displaystyle{%
   \Delta\Gamma_j=\frac{z_j!}{N_j!\big(z_j-N_j\big)!}\,.  %
}
\end{array}
\end{equation}
In the general case, when the number of particles can be small and
take fractional values, the factorials in (29) should be determined
through the gamma function [2], so that the statistical weight (29)
will be written in the form
\begin{equation} \label{30}
\begin{array}{l}
\displaystyle{%
   \Delta\Gamma_j = \frac{\Gamma(z_j+1)}{\Gamma(N_j+1)\Gamma(z_j-N_j+1)}. %
}
\end{array}
\end{equation}
This implies the formula for nonequilibrium entropy
$S=\sum_jS_j=\sum_j\ln\Delta\Gamma_j$, where
\begin{equation} \label{31}
\begin{array}{l}
\displaystyle{%
   S_j\equiv S_{F\!j} = \ln\Gamma(z_j+1) -\ln\Gamma(z_jn_j+1)-\ln\Gamma\big[z_j(1-n_j)+1\big]. %
}
\end{array}
\end{equation}
For a system of fermions in the equations of Section II, which
determine the population of levels and thermodynamic quantities in
the general case, it should be taken
\begin{equation} \label{32}
\begin{array}{l}
\displaystyle{%
   \theta_j(n_j)\equiv\theta_{F\!j}(n_j)\equiv\psi\big[z_j\big(1-n_j\big)+1\big] - \psi\big(z_jn_j+1\big), %
}
\end{array}
\end{equation}\vspace{-5mm}
\begin{equation} \label{33}
\begin{array}{l}
\displaystyle{%
   \theta_j^{(1)}(n_j)\equiv\theta_{F\!j}^{(1)}(n_j)\equiv -\,\psi^{(1)}\big[z_j\big(1-n_j\big)+1\big] - \psi^{(1)}\big(z_jn_j+1\big). %
}
\end{array}
\end{equation}
To calculate these functions one can use the formulas
\begin{equation} \label{34}
\begin{array}{l}
\displaystyle{%
  \theta_{F\!j}(n_j)=\int_0^\infty \frac{e^{-z_jn_jt}-e^{-z_j(1-n_j)t}}{e^t-1}\,dt, \quad %
  \theta_{F\!j}^{(1)}(n_j)=-\int_0^\infty t\,\frac{e^{-z_jn_jt}+e^{-z_j(1-n_j)t}}{e^t-1}\,dt. %
}%
\end{array}
\end{equation}

In the two-level fermion system two cases must be distinguished: A)
the number of particles is less than or equal to the multiplicity of
degeneracy of the lower level\, $0<N\leq z_1$, B) the number of
particles is greater than the multiplicity of degeneracy    
of the lower level $z_1<N\leq z_1+z_2$. In the case A), at zero
temperature all particles are at the ground level $n_1\le 1, n_2=0$,
so that the number of particles, energy and pressure are determined
by the formulas $N=z_1n_1$, $E=\varepsilon_1 N$ and $p=\alpha\big(N/V\big)\varepsilon_1$, %
and the chemical potential $\mu=\varepsilon_1$. The entropy is given
by the formula
\begin{equation} \label{35}
\begin{array}{l}
\displaystyle{%
   S_{F\!A} = \ln\Gamma(z_1+1) -\ln\Gamma(N+1)-\ln\Gamma\big[z_1+1-N\big]. %
}
\end{array}
\end{equation}

If the level is not filled $N<z_1$, then the entropy (35) at $T=0$
is nonzero and the third law of thermodynamics is satisfied in the
Nernst formulation. Only when the level is completely filled $N=z_1$
the entropy turns to zero, and then the third law of thermodynamics
turns out to be valid in the Planck formulation. In the case B) at
zero temperature the lower level is filled, and $N-z_1$ particles
are located on the upper level. Only particles of the upper level
contribute to the entropy, so that
\begin{equation} \label{36}
\begin{array}{l}
\displaystyle{%
   S_{F\!B} = \ln\Gamma(z_2+1) -\ln\Gamma(N+1-z_1)-\ln\Gamma(z_1+z_2+1-N). %
}
\end{array}
\end{equation}
It turns to zero only if both levels are completely occupied
$N=z_1+z_2$. The energy and pressure in this case are given by the
formulas $E=\varepsilon_1z_1+\varepsilon_2(N-z_1)$,
$p=\big(\alpha/V\big)\big[\varepsilon_1z_1+\varepsilon_2(N-z_1)\big]$, %
and the chemical potential $\mu=\varepsilon_2$.

\vspace{0mm} %
\begin{figure}[b!]
\vspace{02mm}  \hspace{0mm}
\includegraphics[width = 7.3cm]{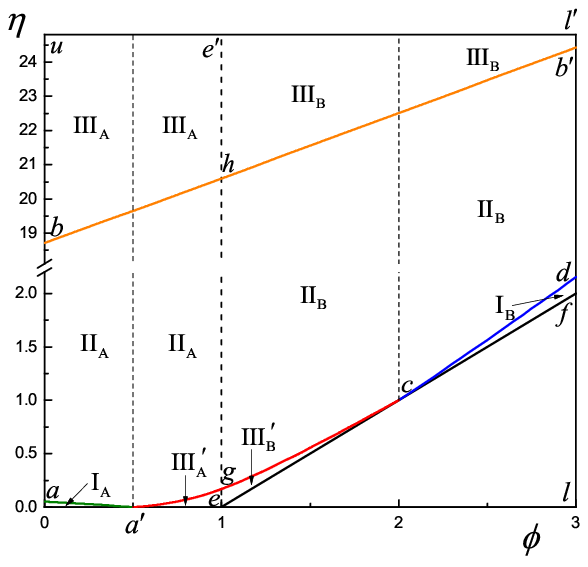} 
\vspace{-4mm} %
\caption{\label{fig04} 
Regions of existence of the characteristic temperatures (37)\,--\,(40): %
${\rm I_A}\,(0aa')$, ${\rm I_B}\,(cdf)$ -- none of these temperatures exist, %
${\rm II_A}\,(aa'ghb)$ -- only $T_{F1}$ (37) exists, %
${\rm II_B}\,(gcdb'h)$ -- only $T_{F3}$ (39) exists, %
${\rm III_A}\,(bhe'u)$ -- $T_{F1}$ (37) and $T_{F2}$ (38) exist,  %
${\rm III_B}\,(hb'l'e')$ -- $T_{F2}$ (38) and $T_{F3}$ (39) exist,  %
${\rm III_A'}\,(a'eg)$ -- $T_{F1}$ (37) and $T_{F4}$ (40) exist,  %
${\rm III_B'}\,(ecg)$ -- $T_{F3}$ (39) and $T_{F4}$ (40) exist.  %
Region $(efl)$ is forbidden, because the condition $N\leq z_1+z_2$
is not satisfied here.
}%
\vspace{-3mm}
\end{figure}

For the case of two-level fermion system four characteristic
temperatures can be determined. At the temperature
\begin{equation} \label{37}
\begin{array}{l}
\displaystyle{%
   T_{F1}=\frac{\Delta\varepsilon}{\Phi_F\big(0,\phi\big)} %
}
\end{array}
\end{equation}
particles from the lower level begin to transit to the upper empty
level. At the temperature
\begin{equation} \label{38}
\begin{array}{l}
\displaystyle{%
   T_{F2}=\frac{\Delta\varepsilon}{\Phi_F\big(\phi\eta^{-1},0\big)} %
}
\end{array}
\end{equation}
all particles from the lower level, which becomes empty, transit to
the upper level. At the temperature
\begin{equation} \label{39}
\begin{array}{l}
\displaystyle{%
   T_{F3}=\frac{\Delta\varepsilon}{\Phi_F\big((\phi-1)\eta^{-1},1\big)} %
}
\end{array}
\end{equation}
particles from the lower, completely filled level, begin to transit
to the upper, partially filled level. Finally, at the temperature
\begin{equation} \label{40}
\begin{array}{l}
\displaystyle{%
   T_{F4}=\frac{\Delta\varepsilon}{\Phi_F\big(1,\phi-\eta\big)} %
}
\end{array}
\end{equation}
particles completely fill the upper level, while some part of them
remains at the lower level. In formulas (37)\,--\,(40) we use the
notation $\Phi_F(n_2,n_1)\equiv \theta_{F2}(n_2)-\theta_{F1}(n_1)$. %
Note that the functions in the denominators of formulas
(37)\,--\,(40), written in the variables $(\phi,\eta)$, depend also
on $N$ as on a parameter. Obviously, the condition of the existence
of these temperatures is the positivity of denominators in
(37)\,--\,(40). These requirements are satisfied not for all values
of the quantities $z_1, z_2, N$. The regions where these
temperatures may exist on the plane $(\phi,\eta)$,\,(27) are shown in Fig.\,4. %

The dependences of the dimensionless temperatures
$\tau_F=T_F\big/\Delta\varepsilon$ on the parameter $\eta$ at a
fixed parameter $\phi$ in different cases are shown in Fig.\,5.
There are four regions of variation of the parameter $\phi$, where
characteristic temperatures depend differently on the parameter $\eta$: %
{\it a}) $0<\phi<1/2$ (Fig.\,5{\it a}), {\it b}) $1/2<\phi<1$ (Fig.\,5{\it b}), %
{\it c}) $1<\phi<2$ (Fig.\,5{\it c}), {\it d}) $\phi>2$ (Fig.\,5{\it d}). %

\begin{figure}[h!]
\vspace{-2mm}  \hspace{0mm}
\includegraphics[width = 14.5cm]{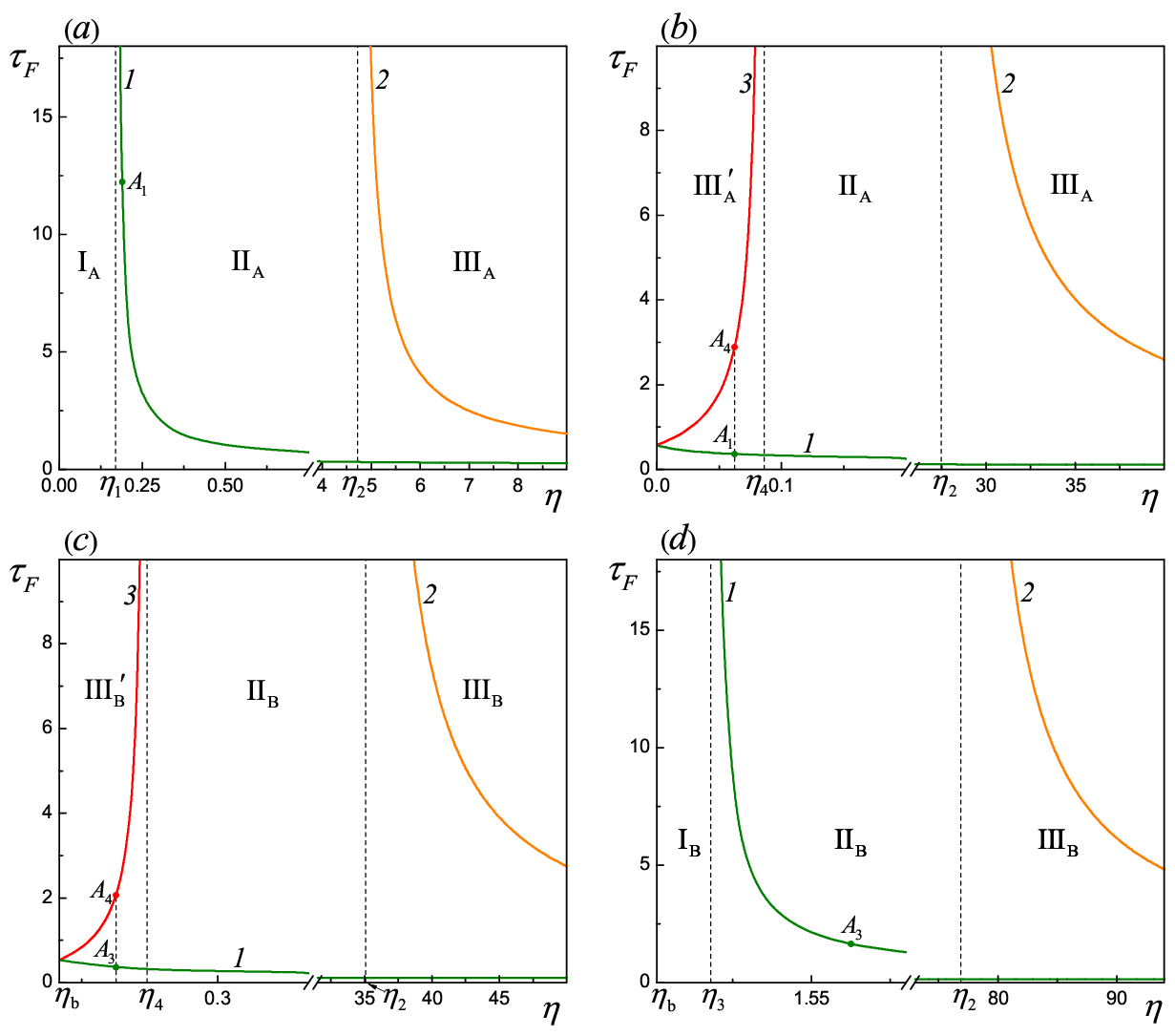} 
\vspace{-4mm} %
\caption{\label{fig05} 
Dependences of the dimensionless temperatures on the parameter $\eta$, %
characterizing the ratio of the degeneracy factors of the upper and lower levels. \newline %
(\!{\it a}) $\tau_{F1}$ (37) -- {\it 1}, $\tau_{F2}$ (38) -- {\it 2}\, %
for $z_1=16$, $N=2$ and $\phi=0.125$; $\eta_1=0.17$, $\eta_2=4.72$; %
$A_1=(0.188, 12.2)$ -- matches to $z_2=3$, $A_2=(4.75, 137.1)$ -- matches to $z_2=76$ (not shown). %
\newline %
(\!{\it b}) $\tau_{F1}$ -- {\it 1}, $\tau_{F2}$ -- {\it 2}, $\tau_{F4}$ (40) -- {\it 3}\, %
for $z_1=16$, $N=14$ and $\phi=0.875$; $\eta_4=0.086$, $\eta_2=27.5$; %
$A_1=(0.063, 0.36)$, $A_4=(0.063, 2.88)$ -- match to $z_2=1$, $A_2=(27.5, 1862.4)$ -- matches to $z_2=440$ (not shown). %
\newline %
(\!{\it c}) $\tau_{F3}$ (39) -- {\it 1}, $\tau_{F2}$ -- {\it 2}, $\tau_{F4}$ -- {\it 3}\, %
for $z_1=16$, $N=18$ and $\phi=1.125$; $\eta_b=0.125$, $\eta_4=0.22$, $\eta_2=35.1$; %
$A_3=(0.186, 0.35)$, $A_4=(0.186, 2.06)$ -- match to $z_2=3$, $A_2=(35.13, 802.1)$ -- matches to $z_2=562$ (not shown). %
\newline %
(\!{\it d}) $\tau_{F3}$ -- {\it 1}, $\tau_{F2}$ -- {\it 2}\, %
for $z_1=16$, $N=40$ and $\phi=2.5$; $\eta_b=1.5$, $\eta_3=1.52$, $\eta_2=76.9$; %
$A_3=(1.56, 1.65)$ -- matches to $z_2=25$, $A_2=(76.9, 13825.9)$ -- matches to $z_2=1230$ (not shown). %
} \vspace{0mm}%
\end{figure}

The temperature dependences of the entropy and heat capacities for
the cases {\it b}) $1/2<\phi<1$ and {\it d}) $\phi>2$ are shown in
Figures 6 and 7.

In the range of the parameter values {\it b}) $1/2<\phi<1$ (Fig.\,6) %
in the region ${\rm III_A}$ (Fig.\,6{\it c}), at temperatures
$\tau<\tau_{F1}$ the entropy $S_{F1}$ is constant and the heat
capacities are equal to zero. At $\tau=\tau_{F1}$ the heat
capacities take on by jumps finite values, then with increasing
temperature they reach maximums and begin to decrease. At
temperature $\tau=\tau_{F2}$ the heat capacities turn to zero by
jumps. The entropy on the interval $\tau_{F1}<\tau<\tau_{F2}$
increases monotonically from $S_{F1}$ to $S_{F2}$. The pressure
increases monotonically with increasing temperature. In the
temperature region with temperature dependences, the populations are
determined by the system of equations (28) with the account of the
substitution $\theta_{B\!j}(n_j)\rightarrow \theta_{F\!j}(n_j)$. %
In the region ${\rm II_A}$ (Fig.\,6{\it b}) there is no the limiting
temperature $\tau_{F2}$, so that at $\tau\rightarrow\infty$ the heat
capacities tend to zero, and the entropy tends to the limiting value
$S_{F\!\infty}$. In the region ${\rm III_A'}$ (Fig.\,6{\it a}), in
the temperature range $\tau_{F1}<\tau<\tau_{F4}$ the heat capacities
decrease monotonically, and at $\tau=\tau_{F4}$ they turn to zero by
jumps. In this case, the entropy on the interval
$\tau_{F1}<\tau<\tau_{F4}$ increases monotonically from $S_{F\!1}$
to $S_{F\!2}$. The variation of the entropy and heat capacities with
temperature in the case {\it c}) $1<\phi<2$ is similar.

\begin{figure}[t!]  
\vspace{-1mm}  \hspace{0mm}
\includegraphics[width = 14.2cm]{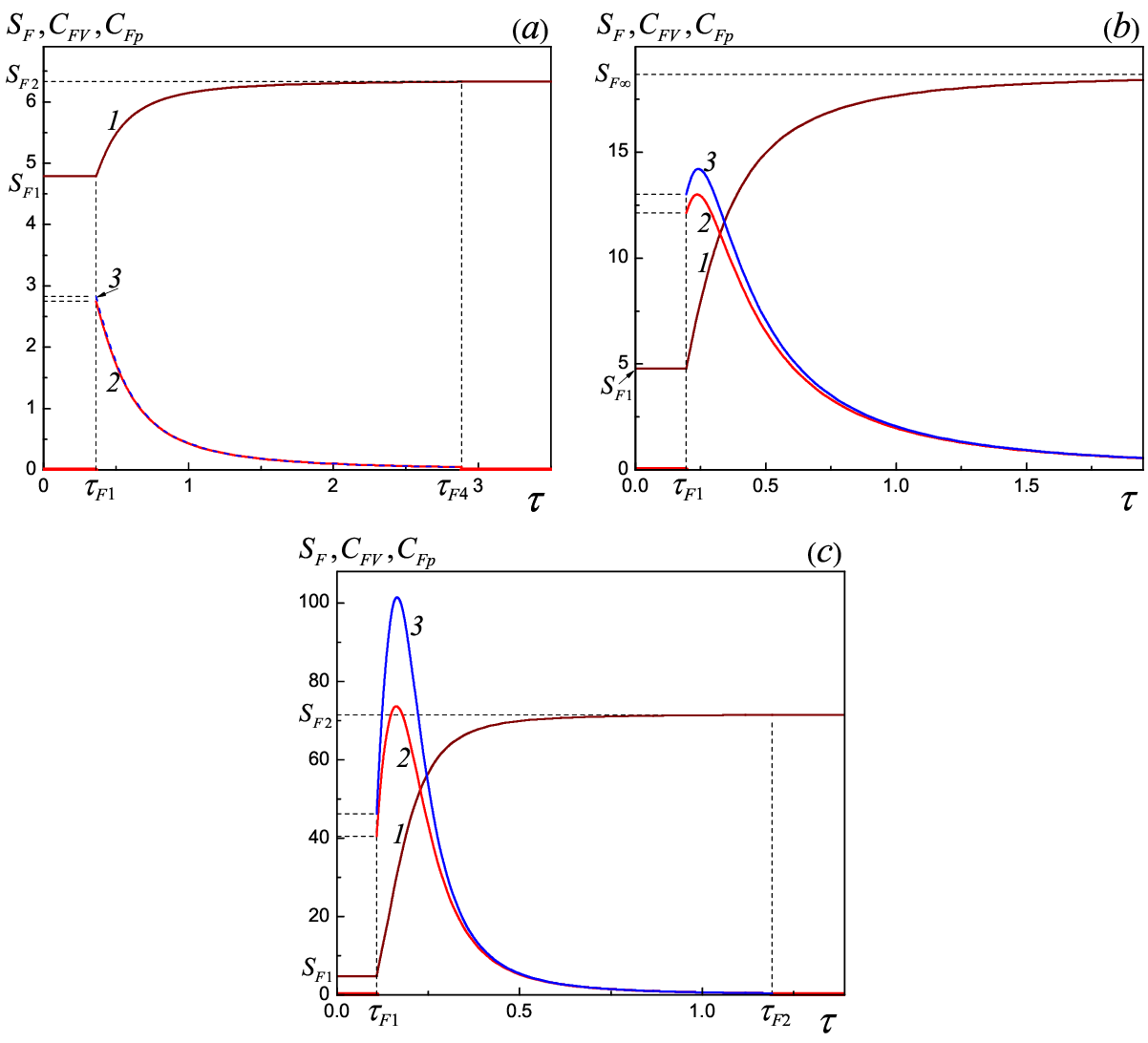} %
\vspace{-4mm} %
\caption{\label{fig05} 
Dependences of the entropy $S_F(\tau)$ ({\it 1}) and the heat
capacities $C_{FV}(\tau)$ ({\it 2}), $C_{Fp}(\tau)$ ({\it 3})
on the interval {\it b}) $1/2<\phi<1$. \newline 
(\!{\it a}) Region ${\rm III_A'}$. %
Jumps of heat capacities at $\tau_{F1}$: $\Delta C_{FV}=2.75$, $\Delta C_{Fp}=2.83$; %
jumps at $\tau_{F4}$: $\Delta C_{FV}=-0.045$, $\Delta C_{Fp}=\Delta C_{FV}-1.6 \cdot 10^{-4}$. %
Entropy values: $S_{F1}=4.8$, $S_{F2}=6.3$. %
Parameters: $z_1=16$, $z_2=1$, $N=14$; $\tau_{F1}=0.36$, $\tau_{F4}=2.88$.  \newline %
(\!{\it b}) Region ${\rm II_A}$. %
Jumps of heat capacities at $\tau_{F1}$: $\Delta C_{FV}=12.14$, $\Delta C_{Fp}=13.02$. %
Entropy values: $S_{F1}=4.8$, $S_{F\infty}=18.7$. %
Parameters: $z_1=16$, $z_2=16$, $N=14$; $\tau_{F1}=0.19$. \newline %
(\!{\it c}) Region ${\rm III_A}$. %
Jumps of heat capacities at $\tau_{F1}$: $\Delta C_{FV}=40.4$, $\Delta C_{Fp}=46.2$; %
jumps at $\tau_{F2}$: $\Delta C_{FV}=-0.397$, $\Delta C_{Fp}=-0.400$. %
Entropy values: $S_{F1}=4.8$, $S_{F2}=71.4$. %
Parameters: $z_1=16$, $z_2=1000$, $N=14$; $\tau_{F1}=0.11$, $\tau_{F2}=1.19$. %
} \vspace{0mm}%
\end{figure}

\begin{figure}[t!]   
\vspace{0mm}  \hspace{0mm}
\includegraphics[width = 14.2cm]{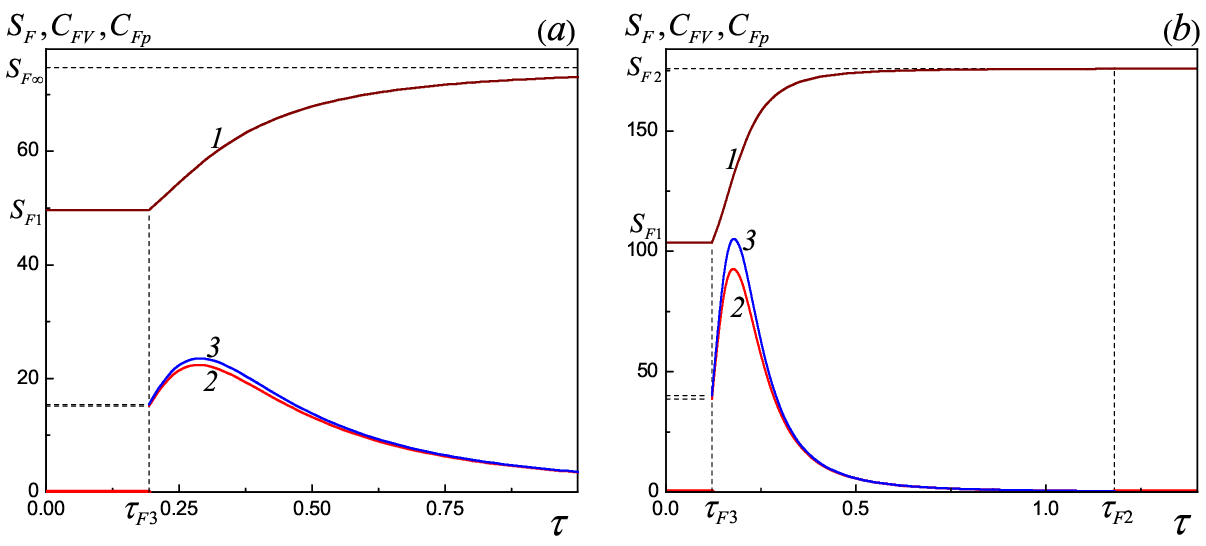} %
\vspace{-4mm} %
\caption{\label{fig07} 
Dependences of the entropy $S_F(\tau)$ ({\it 1}) and the heat capacities 
$C_{FV}(\tau)$ ({\it 2}), $C_{Fp}(\tau)$ ({\it 3}) on the interval {\it d}) $\phi>2$. \newline %
(\!{\it a}) Region ${\rm II_B}$. %
Jumps of heat capacities at $\tau_{F3}$: $\Delta C_{FV}=15.10$, $\Delta C_{Fp}=15.45$. %
Entropy values: $S_{F1}=49.7$, $S_{F\infty}=74.7$. 
Parameters: $z_1=16$, $z_2=128$, $N=34$; $\tau_{F3}=0.19$. 
Maximums on the curves for heat capacities appear at $z_2\simeq 30$. \newline %
(\!{\it b}) Region ${\rm III_B}$. %
Jumps of heat capacities at $\tau_{F3}$: $\Delta C_{FV}=38.57$, $\Delta C_{Fp}=40.01$; %
jumps at $\tau_{F2}$: $\Delta C_{FV}=-0.414$, $\Delta C_{Fp}=-0.415$. %
Entropy values: $S_{F1}=103.6$, $S_{F2}=175.8$. %
Parameters: $z_1=16$, $z_2=2400$, $N=34$; $\tau_{F3}=0.12$, $\tau_{F2}=1.18$. %
} \vspace{0mm}%
\end{figure}

In the range of the parameter values {\it d}) $\phi>2$ (Fig.\,7) in
the region ${\rm I_B}$, the state of the system does not change at
arbitrary temperatures of the thermostat. In the region ${\rm II_B}$
(Fig.\,7{\it a}), at temperatures $\tau<\tau_{F3}$ the entropy
$S_{F1}$ is constant and the heat capacities are equal to zero. At
$\tau_{F3}$ the heat capacities take on finite values by jumps. If
$z_2>30$, then at $\tau>\tau_{F3}$ the heat capacity curves have
maxima and at $\tau\rightarrow\infty$ they tend to zero. For
$z_2<30$ the heat capacities decrease monotonically. The entropy at
$\tau>\tau_{F3}$ monotonically increases to the limiting value
$S_{F\infty}$. In the region ${\rm III_B}$ (Fig.\,7{\it b}) there
exists the limiting temperature $\tau_{F2}$, at which the heat
capacities turn to zero by jumps and the entropy reaches its maximum
value $S_{F2}$. The variation of the entropy and heat capacities
with temperature in the case {\it a}) $0<\phi<1/2$ is similar to the
case {\it d}).

\section{Discussion and conclusions }\vspace{-0mm} %
In this work we studied the thermodynamic properties of systems of
non-interacting bosons and fermions with a small number of
particles. The equations for the average number of particles in each
quantum state for an arbitrary number of particles were previously
obtained by the authors in [2]. In the work, within the framework of
theory [2], the temperature dependences of the entropy, heat
capacities and pressure are calculated under the assumption that
particles can be in two degenerate states.

It is shown that in the case of bosons, which at zero temperature
are all at the lower level, with an increase in temperature,
depending on the multiplicity of degeneracy of the upper level,
three qualitatively different situations are possible. When the
degeneracy factor of the upper level is low $z_2\ll z_1$, all
particles, regardless of the thermostat temperature, remain at the
lower level (region I, in Fig.\,1). The entropy is constant in this
state. Due to the low degeneracy factor of the upper level, the
system cannot receive energy from the thermostat and therefore turns
out to be adiabatically isolated. With a greater degeneracy factor
of the upper level, at a certain temperature $T_{B1}$ (25), the
transition of particles to the upper level becomes possible (region
II, in Fig.\,1). In this case the entropy and pressure begin to
increase with increasing temperature, and the heat capacities at
$T_{B1}$ take on finite values by jumps. At high temperatures in the
limit $T\rightarrow\infty$, particles remain distributed between two
levels with finite populations, the entropy and pressure tend to
constant values, and the heat capacities tend to zero (Fig.\,3{\it a}). %
Finally, with a further increase of the degeneracy factor of the
upper level, a case is possible when at a certain limiting
temperature $T_{B2}$ (26) all particles transit to the upper level,
and the lower level becomes empty (region III, in Fig.\,1). This
temperature can be considered as an analogue of the Bose-Einstein
condensation temperature, below which the filling of the ground
level begins. At $T\ge T_{B2}$ the energy of the system reaches its
maximally possible value, so that with an increase in the thermostat
temperature the transfer of heat from the thermostat to the system
becomes impossible. At $T_{B2}$ the heat capacities turn to zero by
jumps (Fig.\,3{\it b}). Since the entropy at zero temperature for a
system of bosons with account of discreteness of levels turns out to
be finite and non-zero, the third law of thermodynamics is satisfied
in the Nernst formulation.

In a gas of non-interacting fermions at zero temperature there are
several qualitatively different states. If the number of particles
does not exceed the degeneracy factor of the ground level $N\le
z_1$, then all particles at $T=0$ are located at this lower level.
When the degeneracy factor of the upper level is low $z_2\ll z_1$
and $N\le z_1/2$, all particles at any thermostat temperature, as in
the case of bosons, remain at the lower level (region ${\rm I_A}$,
Fig.\,4). In this state the system has a constant entropy and turns
out to be adiabatically isolated. With an increase of the degeneracy
factor of the upper level and for a small number of particles, at a
certain temperature $T_{F\!1}$ (37) the transition of particles to
the upper level becomes possible (region ${\rm II_A}$, Fig.\,4). At
that the entropy and pressure begin to increase, and the heat
capacities take on finite values by jumps. At high temperatures in
the limit $T\rightarrow\infty$, particles remain distributed between
two levels with finite populations, the entropy and pressure tend to
constant values, and the heat capacities tend to zero. With a
further increase of the degeneracy factor of the upper level (region
${\rm III_A}$, Fig.\,4), at a certain limiting temperature
$T_{F\!2}$ (38) all particles transit to the upper level, and the
lower level becomes empty. With an increase in temperature the state
of the system does not change, its entropy and pressure remain
constant, and its heat capacities are equal to zero. When the
condition $N>2z_1$ is satisfied, there is the region ${\rm I_B}$
similar to the region ${\rm I_A}$ (Fig.\,4), where the system
remains in the ground state at all permissible temperatures. Here,
with increasing $z_2$ there also exist states similar to the
previous case $N\le z_1/2$ (regions ${\rm II_B}$ and ${\rm III_B}$,
Fig.\,4). The temperature dependences of the entropy and heat
capacities for the case $N>2z_1$ are shown in Fig.\,7.

If the inequality $z_1/2<N\le z_1$ is satisfied, then even for a low
degeneracy factor of the upper level $z_2\ll z_1$ the region of
adiabaticity of the system is absent at all temperatures, and at
$T_{F\!1}$ (37) the transition of particles to the upper level
begins in it. At the temperature $T_{F4}$ (40) the upper level
proves to be filled, and particles continue to remain at the lower
level (region ${\rm III_A'}$, Fig.\,4). With an increase of the
degeneracy factor of the upper level the system successively passes
into the regions ${\rm II_A}$ and then ${\rm III_A}$, shown in
Fig.\,4. The temperature dependences of the entropy and heat
capacities for this case $z_1/2<N\le z_1$ are presented in Fig.\,6.

In the second qualitatively different situation, when $N>z_1$, at
zero temperature the lower level is completely occupied and $N-z_1$
particles are located at the upper level. When the condition
$z_1<N\le 2z_1$ is satisfied, in the region ${\rm III_B'}$ (Fig.\,4)
at the temperature $T_{F3}$ (39) there begins the transition of
particles to the upper level. At $T_{F4}$ (40) the upper level
proves to be filled, and a part of particles remains at the lower
level. The state does not change with increasing temperature. At
higher values of $z_2$ (region ${\rm II_B}$, Fig.\,4) only the
temperature $T_{F3}$ exists, and in the limit $T\rightarrow\infty$
particles become distributed between two levels with finite
populations. At yet more high values of $z_2$ (region ${\rm III_B}$,
Fig.\,4), in addition to $T_{F3}$ there exists the limiting
temperature $T_{F2}$ (38) at which all particles transit to the
upper level and the lower level becomes empty.

Note that the issue of negative and limiting temperatures in systems
with a limited spectrum was considered by Yu.B. Rumer [6,\,7]. In
our case, by the temperature of a system with a small number of
particles we mean the temperature of the thermostat with which it is
in equilibrium, so that the temperature is always positive. In the
above consideration, due to the finite number of levels, there exist
limiting temperatures at which the energy of the system becomes
maximum, so that a further increase in the temperature of the
thermostat does not lead to an increase in the energy of the system.

In conclusion, we formulate the general features of the
thermodynamics of two-level systems with a finite number of bosons
and fermions:

1.\,\,At zero temperature the entropy can be non-zero, so that the
third law of thermodynamics is satisfied in the Nernst formulation.

2.\,\,With a small degeneracy factor of the upper level particles
can remain at the lower level at arbitrary temperatures.

3.\,\,As the degeneracy factor of the upper level increases, the
transition of particles from the lower to the upper level becomes
possible, such that in the limit of high temperatures particles are
distributed between two levels with finite populations.

4.\,\,At yet greater degeneracy factor of the upper level there
exists the limiting temperature at which the upper level becomes
maximally filled, so that the energy reaches its greatest value. A
further increase in the thermostat temperature does not change the
state of the system.


\end{document}